\documentclass[aps,prb,twocolumn,floatfix,notitlepage, superscriptaddress,longbibliography]{revtex4-2}
\usepackage[utf8]{inputenc}
\usepackage{amsfonts,amsmath,amssymb,graphicx,epstopdf,verbatim}
\usepackage{rotating}
\usepackage{epstopdf}
\usepackage{xcolor}
\usepackage{natbib}
\usepackage[colorlinks]{hyperref}
\definecolor{darkred}{rgb}{0.5,0,0}
\definecolor{darkgreen}{rgb}{0,0.5,0}
\definecolor{darkblue}{rgb}{0,0,0.5}
\hypersetup{colorlinks,
linkcolor=darkred,
filecolor=darkgreen,
urlcolor=darkblue,
citecolor=darkgreen}
\usepackage{braket}
\usepackage{overpic}

\newcommand{\Tr}[1]{\mathrm{Tr}\!\left[#1\right]}

\newcommand{\calN}{\mathcal{N}}

\newcommand{\ud}{\mathrm{d}}
\newcommand{\nep}{\operatorname{e}}

\newcommand{\opcdag}[1]{{\hat{c}^{\dagger}}_{#1}}

\newcommand{\opc}[1]{{\hat{c}^{\phantom \dagger}}_{#1}}

\begin{document}

%\title{Quantum state diffusion, entanglement transitions and quantum bifurcations with long-range dissipators}
\title{Entanglement transitions and quantum bifurcations\\ under continuous long-range monitoring}

\author{Angelo Russomanno}
\affiliation{Scuola Superiore Meridionale, Università di Napoli Federico II
Largo San Marcellino 10, I-80138 Napoli, Italy}
\affiliation{Dipartimento di Fisica ``E. Pancini'', Università di Napoli Federico II, Complesso di Monte S. Angelo, via Cinthia, I-80126 Napoli, Italy}

\author{Giulia Piccitto}
\affiliation{Dipartimento di Fisica dell’Università di Pisa and INFN, 
Largo Pontecorvo 3, I-56127 Pisa, Italy}
\affiliation{Dipartimento di Matematica e Informatica, Università di Catania, Viale Andrea Doria 6, 95125, Catania, Italy}

\author{Davide Rossini}
\affiliation{Dipartimento di Fisica dell’Università di Pisa and INFN, 
Largo Pontecorvo 3, I-56127 Pisa, Italy}

\begin{abstract}
  We study the asymptotic bipartite entanglement entropy of the quantum trajectories of a free-fermionic system,
  when subject to a continuous nonlocal monitoring. The measurements are described by Gaussian-preserving
  two-point operators, whose strength decays as a power-law with exponent $\alpha$.
  Different behaviors of the entanglement entropy with the system size emerge: for $\alpha$ below a given threshold value 
  a volume-law behavior sets in, while for larger $\alpha$ we observe a transition from subvolume to area-law,
  whose exact location depends on the measurements rate and on the presence of a Hamiltonian dynamics.
  We also consider the expectation probability distribution of the measurement operators, 
  and find that this distribution features a transition from a unimodal to a bimodal shape. We discuss the possible connections
  between this qualitative change of the distribution and the entanglement transition points.
\end{abstract}

\maketitle

\section{Introduction}
Nowadays it is widely believed that entanglement, alias a kind of quantum correlations with no classical
analog~\cite{Nielsen, RevModPhys.81.865}, plays an important role in the equilibrium and the out-of-equilibrium
physics of quantum many-body systems~\cite{Amico_RMP}. 
A prototypical example is that of the entanglement entropy for a pure state, which is defined
as the von Neumann entropy of the reduced density matrix of a given portion of the full system. %~\cite{Nielsen}.
Due to its peculiar scaling properties at the critical point~\cite{Vidal2003, Vidal2003b}, it may act as
a witness of the presence of quantum phase transitions.
Moreover, in nonequilibrium conditions, it generally increases linearly in time to eventually attain
an asymptotic value proportional to the system size~\cite{Alba_2017, Alba_2018, Singh_2016},
and the slope of this scaling contains information on the thermalization properties of the system~\cite{PhysRevB.102.144302}. 
Such scenario changes qualitatively in the presence of disorder: For example, in many-body localized phases,
the entanglement entropy undergoes a characteristic, much slower, logarithmic increase
in time (see Ref.~\cite{Abanin_RMP} for a review).

More recently, the focus has been moved to situations beyond the unitary dynamics, which consider
the evolution of monitored systems. The interplay between the intrinsic dynamics of the system and that induced
by the quantum measurement process can lead to a variety of scaling regimes for the asymptotic entanglement entropy,
giving rise to the so called entanglement transitions. 
In this framework, an extensive number of works has been focusing on local measurements (either discrete or continuous in time)
performed in monitored quantum 
circuits~\cite{Li2018, Chan2019, Skinner2019, Szynieszewski2019, Vasseur2021, Bao2021, Nahum2020, Chen2020,Li2019, Jian2020, Li2021, Szyniszewski2020, Turkeshi2020, Lunt2021, Sierant2022_B, Nahum2021, Zabalo2020, Sierant2022_A, Chiriaco2023, Klocke2023},
as well as non-interacting~\cite{DeLuca2019,Nahum2020, Buchhold2021,Jian2022, Coppola2022, Fava2023, Poboiko2023, Jian2023, Merritt2023, Alberton2021, Turkeshi2021, Szynieszewski2022, Turkeshi2022, Piccitto2022, Piccitto2022e, Tirrito2022, Paviglianiti2023}
and interacting~\cite{Lunt2020,Rossini2020, Tang2020, Fuji2020, Sierant2021, Doggen2022, Altland2022} Hamiltonian systems.
Moreover, there exists a deep connection between measurement-induced phases and the encoding/decoding properties of a quantum channel~\cite{Gullans2020_A, Gullans2020_B, Loio2023, Choi2020, Bao2020, Bao2021_A,Fidkowski2021, Bao2021_B, Barratt2022_A,Dehgani2023, Kelly2022}.
Situations where the dynamics is only induced by random measurements of non-local
string operators (measurement-only dynamics) have been also considered, finding different scaling regimes of the entanglement entropy, according to the statistics of the randomly measured operators, and the range and the nature of the strings~\cite{Ippoliti2021, Sriram2022}.

Among the various theoretical models of monitored quantum systems, considerable coverage has been dedicated to the dynamics
of fermionic Gaussian states, in the presence of quadratic Hamiltonians and Gaussian-preserving measurement
processes (see, e.g., Refs.~\cite{DeLuca2019, Lang2020, Alberton2021, Turkeshi2021, Turkeshi2022, Piccitto2022, Piccitto2022e, Coppola2022, Tirrito2022, Minato2022, Szynieszewski2022, Zerba2023, Paviglianiti2023, Poboiko2023}),
as they are amenable to an accurate numerical treatment up to relatively large sizes.
In this framework, for short-range Hamiltonians and local measurements, area-law (saturation to a finite value)
or logarithmic scaling of the asymptotic entanglement entropy with the system size have been reported.
A somewhat richer situation has been found for Hamiltonians with extended power-law interactions,
although keeping the measurement operators onsite, where regimes with a power-law scaling of the entanglement entropy
with the system size are possible~\cite{Minato2022, M_ller_2022}.
Something similar has been considered in the context of quantum circuits~\cite{Block_2021, Sharma_2022}. 
In a recent paper, we have also shown that the measurement-only dynamics through operators connecting two distant sites 
can give rise to a non-trivial entanglement entropy dynamics, with a fast growth of the entanglement entropy~\cite{Piccitto2023}. 

In this work we deal with the quantum dynamics of a Kitaev chain under continuous nonlocal monitoring, which
can be cast as a quantum state diffusion unraveling~\cite{Gisin1992,Jacobs2014_Book}
of a Lindblad master equation with long-range Lindblad operators.
Specifically, we consider two-point fermionic measurement operators, suitably chosen
to preserve Gaussianity, where the coupling decays as a power-law with some exponent $\alpha>0$.
In the context of dissipation engineering, similar kind of dissipators have been already scrutinized
in some recent
works~\cite{PhysRevResearch.4.013089,PhysRevB.105.184305,PhysRevLett.129.050603,PhysRevB.106.224308};
these can be realized with two-level atoms in lossy cavity QED experiments, using a magnetic field gradient and a
Raman drive with multiple sidebands~\cite{PhysRevResearch.4.013089}. 
In noninteracting spins monitored by infinite-range operators, an entanglement transition
from area-law to sublogarithm scaling can occur~\cite{passarelli2023postselectionfree}.

Here we first consider the asymptotic bipartite entanglement entropy and find a rather rich phenomenology:
For $\alpha$ smaller than a threshold value $\alpha^\star_1$, it obeys a volume law,
suggesting a strong entangling power of the long-range measurement operators.
For intermediate values of $\alpha$, a crossover region emerges, in which the entanglement entropy
scales non-trivially with the size.
For $\alpha$ larger than $\alpha^\star_2$ $(> \alpha^\star_1)$, we recover the area-law scaling
observed in the presence of onsite measurements.
The fact that $0.5 \lesssim \alpha^*_1 \lesssim 1$, independently of the Hamiltonian parameters and
of the coupling with the measurement apparatus, is suggestive.
Indeed, $\alpha=1$ corresponds to the threshold below which both the unitary (Hamiltonian) long-range dynamics
in one dimension~\cite{Ruffo, PhysRevLett.120.130601} and at least a single case of Lindblad long-range dynamics
in one dimension~\cite{PhysRevB.106.224308} are exactly described by the mean-field approximation.

We also focus on a measurement-only dynamics, i.e., such that there is no Hamiltonian providing
a unitary part in the evolution.
In that case, we still have evidence that $0.5 \lesssim \alpha^*_1 \lesssim 1$. Besides that,
we can locate the transition point between subvolume and area-law behavior at $\alpha_2^* \sim 2$,
suggesting an even more interesting comparison with the behavior of long-range Hamiltonians,
where the system behaves short-range above the threshold $\alpha=2$~\cite{Ruffo,PhysRevLett.120.130601}.

Finally, we study the expectation probability distribution of the measurement operators.
When increasing $\alpha$, the distribution over a single quantum trajectory of the expectation values
of such operators undergoes a transition from unimodal (one maximum) to bimodal (two maxima), at a point
$\bar{\alpha}$ that is not immediately related with the change of scaling for the entanglement entropy.
Such transition is reminiscent of the bifurcations occurring in nonlinear driven-dissipative classical dynamical systems, where a single stable stationary point splits into two~\cite{strogatz:book,cross:book}.
{Here, due to the presence of quantum fluctuations and classical noise, there are no stationary points, and their equivalent are }the maxima of the distribution, that move from being one to two.
In view of this analogy, we dub the unimodal-bimodal transition
of the distribution of the expectations as a ``quantum bifurcation''.

The paper is organized as follows. In Sec.~\ref{mod:sec}, we define our model, specifying both the Hamiltonian
and the measurement operators, together with the bipartite entanglement entropy we are going to analyze.
In Sec.~\ref{qsd:sec}, we introduce the quantum state diffusion unraveling of the Lindblad master equation
and discuss how to treat the time evolution of the system, preserving the Gaussian form of the wavefunction. 
Section~\ref{res:sec} is devoted to the presentation of our numerical findings summarized above,
for the entanglement entropy (Sec.~\ref{ent:sec}) and for the expectation distribution of the measurement operators
(Sec.~\ref{dist:sec}). In Sec.~\ref{conc:sec} we draw our conclusions.

%.......................................................................................................................................%

\section{Model}
\label{mod:sec}

We start from a system of spinless fermions on a one-dimensional lattice with $N$ sites, described by the Kitaev Hamiltonian~\cite{Kitaev_2001}
\begin{equation}
  \label{hami:eqn}
  \hat{H} = \sum_i \Big[ J \big( \hat{c}_i-\hat{c}_i^\dagger \big) \big( \hat{c}_{i+1}+\hat{c}_{i+1}^\dagger \big) +2h \, \opcdag{i}\opc{i} \Big]\,.
\end{equation}
The real constants $J$ and $h$ stand for, respectively, the nearest-neighbor coupling and the chemical potential
$\mu \equiv 2h$, while $\hat c_{i}^{(\dagger)}$ are annihilation (creation) operators on the $i$th site ($i=1,\ldots, N$),
exhibiting canonical anticommutation relations.
The Hamiltonian~\eqref{hami:eqn} is responsible for the unitary part of the dynamics.
We notice that this model can be mapped, via a Jordan-Wigner transformation, onto a
quantum Ising chain in a transverse field~\cite{Sachdev,PFEUTY197079}. Hereafter we set $J=1$ as a energy scale
and work in units of $\hbar=1$.

We consider the Lindblad master equation 
\begin{equation}
  \label{eq:Master}
  \frac{d}{d t} \rho(t) = - i \big[ \hat H, \rho(t) \big]
  + \frac{\gamma}{2} \sum_i \Big( \{ {\hat \ell}_i^\dagger {\hat \ell}_i, \rho \} - 2 \, {\hat \ell}_i \rho \, {\hat \ell}_i^\dagger \Big)
\end{equation}
with measurement operators 
\begin{equation}
  \label{li:eqn}
  \hat{\ell}_i = \sum_{j}f_{i\,j} \big( \hat{c}_i-\hat{c}_i^\dagger \big) \big( \hat{c}_j+\hat{c}_j^\dagger \big) \,,
\end{equation}
and focus on its quantum state diffusion unraveling.
This corresponds to a continuous time monitoring of the system, which is described by
the following stochastic Schr\"odinger equation for the pure state $\ket{\psi_t}$:
\begin{align}
  \label{stoca:eqn}
  \ud\ket{\psi_t} = &-i \hat{H}\ud t\ket{\psi_t} + \sum_i \Big(\sqrt{\gamma}[\hat{\ell}_i-\braket{\hat{\ell}_i}_t]\ud W_t^i\nonumber\\
  &-\frac{\gamma}{2}[\hat{\ell}_i-\braket{\hat{\ell}_i}_t]^2\ud t\Big)\ket{\psi_t}\,,
\end{align}
where $\gamma>0$ is the coupling strength with the measurement apparatus,
$\braket{\hat{\ell}_i}_t = \bra{\psi_t} \hat \ell_i \ket{\psi_t}$, and $W_t^i$ are independent Wiener processes
describing a quantum state diffusion process that unravel the equation~\eqref{eq:Master}.

In Eq.~\eqref{li:eqn}, the real prefactor $f_{i\,j}$ is assumed to algebraically decay with the distance $D_{i,j}$ between site $i$ and site $j$,
such that
\begin{equation}\label{fiji:eqn}
  f_{i\,j} = \frac{1}{N(\alpha)} \, \frac{1}{(1+D_{i,j})^\alpha}\, , \quad\; (\alpha \geq 0) \,.
\end{equation}
Here $N(\alpha) = (N-1)^{-1} \sum_{i, j} (1+D_{i,j})^{-\alpha}$ is a proper normalization constant (the Kac factor),
ensuring extensivity in the system~\cite{kac}. In what follows, we choose periodic boundary conditions for fermions (such that
$\hat c_{j+N}^{(\dagger)} \equiv \hat c_{j}^{(\dagger)}$, for any $j>N$), consequently $D_{i,j} = \min(|i-j|,N-|i-j|)$. Notice also that
the $\hat{\ell}_i$ are Hermitian, ${\hat \ell}_i =\hat \ell_i^\dagger$.

An important property of the operators ${\hat \ell}_i$ is that
\begin{equation}
  \label{quaquad:eqn}
  \hat{\ell}_i^2 = \sum_{j, l} f_{i\,j} f_{i\,l} \big (\hat{c}_j+\hat{c}_j^\dagger \big)
  \big( \hat{c}_l+\hat{c}_l^\dagger \big) = \sum_{j} f_{i\,j}^2\,,
\end{equation}
where we used the anticommutation relations for fermions and the fact that $(\hat c_i - \hat c_i^\dagger)^2 = -1$.
Thanks to this property, Eq.~\eqref{stoca:eqn} can be seen as a Schr\"odinger equation
with a non-Hermitian quadratic Hamiltonian. As a consequence, the state $\ket{\psi_t}$ keeps a simple Gaussian form,
described by just $N(N-1)/2$ complex independent parameters, as we better discuss in Sec.~\ref{qsd:sec}.
One can thus push the numerics to system sizes of some hundreds of sites and investigate how the presence
of power-law decaying measurement operators affects the production of entanglement during the quantum dynamics.

To this purpose, we concentrate on the entanglement entropy of a subchain
of length $l$, averaged over different quantum trajectories
\begin{equation}
  \overline{S_l (t)} \equiv - \overline{\Tr{\rho_l \ln \rho_l}}\,,
  \label{Eq:S}
\end{equation}
where the logarithm is taken in the natural basis.
Here, $\rho_l(t) = \text{Tr}_{N-l}\big[\!\ket{\psi_t}\bra{\psi_t}\!\big]$ is the reduced density matrix
of the subchain and $\ket{\psi_t}$ is the (pure) state of a single quantum trajectory
given by a single realization of the stochastic Schr\"odinger equation dynamics in Eq.~\eqref{stoca:eqn}
(see also Sec~\ref{qsd:sec}).

To obtain the average entanglement entropy, we evaluate it on each single stochastic quantum trajectory
and then ensemble-average over different realizations.
In our analysis, we will mostly focus on the asymptotic long-time value 
\begin{equation}
  S_l = \lim_{T\to \infty}\int_{t^*}^T dt' \, \overline{S_l(t')}\,.
  \label{Eq:S_inf}
\end{equation}
As discussed in Ref.~\cite{glen}, for fermionic Gaussian states, the entanglement entropy can be determined
from the knowledge of the correlation functions, that are introduced in the next section.

%.......................................................................................................................................%

\section{Dynamics under continuous monitoring}
\label{qsd:sec}

Equation~\eqref{stoca:eqn} can be discretized in time and cast as a sequence of Trotterized evolution steps
that, in the limit $\Delta t~\to~0$, converge back to Eq.~\eqref{stoca:eqn}~\cite{DeLuca2019}. In each Trotterized step,
the measurement and the unitary part of the dynamics act separately and in sequence:
\begin{equation}
  \label{eq:SSE}
  \ket{\psi_{t+\Delta t}} \simeq C \nep^{\sum_i (A_i \hat{\ell}_i - \gamma \hat{\ell}_i^2 \Delta t )}
  \nep^{-i\hat{H}\Delta t} \ket{\psi_t} \,,
\end{equation}
where we have defined
\begin{equation}
  \label{eq:Ai_def}
  A_i \equiv \sqrt{\gamma} \, \Delta W_t^i + 2 \gamma \braket{ {\hat \ell}_i}_t\Delta t,
\end{equation}
with $\Delta W_t^i$ being independent real Gaussian
random variables with vanishing expectation value and variance $\Delta t$.

Expression~\eqref{eq:SSE} can be further simplified by using Eq.~\eqref{quaquad:eqn}. 
In this way one can rewrite Eq.~\eqref{eq:SSE} in the simpler form
\begin{equation}
  \label{eq:SSE2}
  \ket{\psi_{t+\Delta t}}\simeq \tilde{C} \nep^{\sum_i A_i\hat{\ell}_i} \nep^{-i\hat{H}\Delta t} \ket{\psi_t} \,,
\end{equation}
where the irrelevant constant $\exp \big( -\gamma \, \Delta t \, \sum_{j} f_{i\,j}^2 \big)$, coming from the exponential
of ${\hat \ell}_i^2$, has been absorbed into the normalization prefactor $\tilde C$.

Being both $\hat{\ell}_i$ and the Kitaev Hamiltonian quadratic in the fermionic operators $\hat c_j^{(\dagger)}$,
when starting from an initial Gaussian state, the time evolution of Eq.~\eqref{eq:SSE2} preserves
Gaussianity. In particular, the state $\ket{\psi_t}$ can be cast as
\begin{equation}
  \label{stato:eqn}
  \ket{\psi_t} = {\calN}_t \, \exp{\bigg(\frac{1}{2} \sum_{j_1,j_2} \big[ {\bf Z}_t \big]_{j_1,j_2} \opcdag{j_1} \opcdag{j_2}\bigg) }
  \, |0\rangle ,
\end{equation}
where $\ket{0}$ denotes the vacuum state of the ${\hat c}$-fermions,
and is thus uniquely described by the $N\times N$ antisymmetric matrix ${\bf Z}_t$ (being ${\bf Z}_t$ antisymmetric, it is described by $N(N-1)/2$ complex parameters).
From the matrix ${\bf Z}_t$, one can easily derive any two-point correlation functions.
Defining
\begin{equation}
  \big[ {\bf G}_t \big]_{j,l} \equiv \braket{\psi_t|\opcdag{l} \opc{j}|\psi_t}, \quad
  \big[ {\bf F}_t \big]_{j,l} \equiv \braket{\psi_t|\opc{l} \opc{j}|\psi_t}\,,
\end{equation}
the correlation matrices can be written in terms of the matrix ${\bf Z}_t$ as~\cite{PhysRevB.93.224431}
\begin{equation}
  \label{corr_mat:eqn}
  {\bf G}_t = \big( \boldsymbol{1}+{\bf Z}_t\,{\bf Z}_t^\dagger \big)^{-1}{\bf Z}_t\,{\bf Z}_t^\dagger, \quad
  {\bf F}_t = \big( \boldsymbol{1}+{\bf Z}_t\,{\bf Z}_t^\dagger \big)^{-1}{\bf Z}_t\,,
\end{equation}
where $\boldsymbol{1}$ is the $N \times N$ identity matrix.
Being ${\bf Z}_t^T = -{\bf Z}_t$, we see that ${\bf G}_t = {\bf G}_t^T$ and ${\bf F}_t = - {\bf F}_t^T$.

In the next subsection we show a simple numerical prescription (whose computational requirements scale
polynomially with $N$) to evaluate the matrix ${\bf Z}_t$ after the application of the unitary
and the dissipative part of the evolution step in Eq.~\eqref{eq:SSE2} (and, therefore, the entanglement entropy).

%---------------------------------------------------------------------------------------------------------------------------------------%
\subsection{Evolution of the matrix ${\bf Z}_t$}
\label{sec:Zevol}

It is possible to write a system of ordinary differential equations for the matrix ${\bf Z}_t$, describing
the evolution in Eq.~\eqref{stato:eqn} and efficiently solvable (an alternative derivation
can be found in Ref.~\cite{Piccitto2023}).

Both the action of the unitary step and the measurement step in Eq.~\eqref{eq:SSE2} can be described
as the application, to a Gaussian state of the form Eq.~\eqref{stato:eqn}, of an operator
of the form $\nep^{-\xi \hat T \Delta}$, where 
\begin{equation}
  \hat T = \sum_{i,j} \big( {\bf D}_{i,j} \, \hat c_i^\dagger \hat c_j + {\bf O}_{i,j} \, \hat c_i^\dagger \hat c_j^\dagger
  + \text{h.c.} \big)
\end{equation}
is a generic (Hermitian) quadratic operator, $\Delta$ is real, and $\xi = \{i, -1\}$ accounts for a dynamics
in real or in imaginary time, respectively.
Let us define $\ket{\psi}$ as a Gaussian state of the form~\eqref{stato:eqn},
  to which we apply the operator $\nep^{-\xi \hat T \Delta}$, and ${\bf Z}$
  the corresponding antisymmetric matrix.
After this operation the state
$\ket{\psi'} \equiv \nep^{-\xi \hat T \Delta}\ket{\psi}$ [being $\ket{\psi}$ a generic Gaussian state described
  by the matrix $\mathbf{Z}$, as in Eq.~\eqref{stato:eqn}] is still Gaussian, and its corresponding
$\mathbf{Z}'$ matrix is obtained by integrating the system of ordinary differential equations 
\begin{equation}
  \label{stuni:eqn}
  \xi\frac{\ud}{\ud s}{\bf Z}(s)
  = 2 \big[ {\bf D}\cdot{\bf Z}(s) + {\bf Z}(s)\cdot{\bf D}+{\bf O}+{\bf Z}(s)\cdot{\bf O}\cdot{\bf Z}(s) \big] \,.
\end{equation}
from $s=0$ to $\Delta$, with initial conditions ${\bf Z}(0) \equiv {\bf Z}$, as shown in Ref.~\cite{PhysRevB.93.224431}.

The unitary step of Eq.~\eqref{eq:SSE2} is obtained by posing $\xi=i$, $\Delta = \Delta t$,
and $\hat T = \hat H$, such that
\begin{equation}
  \begin{aligned} 
    & {\bf D}_{i,i+1} = \phantom{-}{\bf D}_{i,i-1}  = -J/2, \quad {\bf D}_{i,i} = h \, , \\ 
    & {\bf O}_{i,i+1} = -{\bf O}_{i,i-1} = -J/2 \,,
  \end{aligned}
\end{equation}
and zero otherwise.
Analogously, the dissipative step can be obtained by posing $\xi=-1$, $\Delta=1$, and $\hat T = \sum_i A_i \hat \ell_i$
[$A_i$ are the real coefficients defined in Eq.~\eqref{eq:Ai_def}].
Using the anticommutation relations and the symmetry of the couplings $f_{i\,j} = f_{j\,i}$, we can write 
\begin{align}
  & \sum_i A_i \hat{\ell}_i = \sum_{i,j} \Big[ A_i f_{i\,j} \big(\hat{c}_i\hat{c}_j+\hat{c}_j^\dagger\hat{c}_i\big) +\text{h.c.} \Big] \\
  & = \frac{1}{2} \sum_{i,j} \Big[(A_i-A_j) f_{i\,j}\hat{c}_i\hat{c}_j\!+ (A_i + A_j)\hat{c}_j^\dagger\hat{c}_i + \text{h.c.} \Big], \nonumber
\end{align}
so that in Eq.~\eqref{stuni:eqn} one has
\begin{equation}
  \label{AA:eqn}
        {\bf D}_{i,j} = -\tfrac{1}{2}(A_i+A_j) f_{i\,j} \, \quad
        {\bf O}_{i,j} = -\tfrac{1}{2}(A_i-A_j) f_{i\,j}\,.
\end{equation}

Defining $N\times N$ matrices ${\bf U}(s)$, ${\bf V}(s)$ such that
\begin{equation}
  \label{eq:UVbogo}
  {\bf U}^\dagger(s) \, {\bf Z}(s) = -{\bf V}^\dagger(s) \,,
\end{equation}
we can show~\cite{nota1} that, if and only if ${\bf Z}(s)$ obeys Eq.~\eqref{stuni:eqn},
then ${\bf U}(s)$ and ${\bf V}(s)$ satisfy the linear system of differential equations
  \begin{equation}
    \label{stdui:eqn}
    \xi\frac{\ud}{\ud s}\left(\begin{array}{c}{\bf U}(s)\\{\bf V}(s)\end{array}\right) = \left(\begin{array}{cc}\phantom{-}{\bf D}&\phantom{-}{\bf O}\\-{\bf O}&-{\bf D}\end{array}\right)\left(\begin{array}{c}{\bf U}(s)\\{\bf V}(s)\end{array}\right)\,.
\end{equation}
This can be straightforwardly integrated to give~\cite{glen, Piccitto2023}
\begin{equation}
  \label{stuni1:eqn}
  \left(\begin{array}{c} {\bf U'}\\{\bf V'}\end{array}\right) =
	  \exp\left[-2 \xi \Delta \left(\begin{array}{cc}\phantom{-}{\bf D}&\phantom{-} {\bf O}\\-{\bf O}&-{\bf D}\end{array}\right)\right]\left(\begin{array}{c} {\bf U}(0)\\{\bf V}(0)\end{array}\right)\,,
\end{equation}
where ${\bf U}(0)$ and ${\bf V}(0)$ correspond to the initial condition ${\bf Z}(0)$ for $\ket{\psi}$.

The above observation provides a direct and simple solution to the problem of finding how the matrix
${\bf Z}$ of a Gaussian state, as that in Eq.~\eqref{stato:eqn}, is modified after the evolution~\eqref{eq:SSE2}.
The latter is composed of a unitary step, followed by a dissipative step: any time a given operator
of the form $e^{- \xi \hat T \Delta}$ 
is applied to the state~\eqref{stato:eqn}, the matrix ${\bf Z} \equiv - [{\bf U}^\dagger]^{-1} \, {\bf V}^\dagger$
is transformed into ${\bf Z}' \equiv - [{\bf U}'^\dagger]^{-1} \, {\bf V}'^\dagger$,
the matrices ${\bf U}'$ and ${\bf V}'$ being expressed as in Eq.~\eqref{stuni1:eqn}.

In the measurement step, to restore the normalization of the state it is necessary to perform the QR decomposition
\begin{equation}
  \left(\begin{array}{c} {\bf U'}\\{\bf V'}\end{array}\right) =
  \left(\begin{array}{c} {\bf U_Q}\\{\bf V_Q}\end{array}\right) {\bf R} \,,
\end{equation}
where ${\bf R}$ is a $L\times L$ upper triangular matrix and ${\bf U_Q}$ and ${\bf V}_Q$ obey the unitarity condition ${\bf U}_Q^\dagger {\bf U}_Q + {\bf V}_Q^\dagger {\bf V}_Q = \boldsymbol{1}$. From the one side, the QR decomposition does not modify the matrix ${\bf Z}$ that defines the state, since it is easy to check that ${\bf Z'} = - [{\bf U}'^\dagger]^{-1} \, {\bf V}'^\dagger = - [{\bf U}_Q^\dagger]^{-1} \, {\bf V}_Q^\dagger$. From the other side, it restores unitarity~\cite{DeLuca2019, Piccitto2023}, allowing the evaluation of the correlation matrices as
\begin{equation}
   {\bf G}'=\boldsymbol{1}-{\bf U}_Q{\bf U}_Q^\dagger, \quad {\bf F}'=- {\bf U}_Q{\bf V}_Q^\dagger\,,
\end{equation}
as one can easily check by substituting ${\bf Z'} = - [{\bf U}_Q^\dagger]^{-1} \, {\bf V}_Q^\dagger$
in Eqs.~\eqref{corr_mat:eqn} and by imposing the unitarity condition.

%.......................................................................................................................................%

\section{Results}
\label{res:sec}

The results presented below have been obtained by initializing the system in the ground state
of the Hamiltonian~\eqref{hami:eqn} with $J=1$, $h_i=100$,
and letting it evolve after a sudden quench of the field to $h=0.5$.
We checked that the asymptotic value of the entanglement entropy, as well as the expectation probability
distribution of the measurement operators, are not affected by the choice of $h_i$ and weakly depend on $h$.
Therefore, without loss of generality, hereafter we keep them fixed.

To compute the entanglement entropy, we choose a balanced bipartition by taking $l = N/2$ [see Eq.~\eqref{Eq:S}],
and finally perform the averages in Eqs.~\eqref{Eq:S} and~\eqref{Eq:S_inf}
over a given number $N_{\rm r}$ of realizations of the stochastic process.
On the other hand, to obtain the full counting statistics, we evolve a single quantum trajectory
up to a long time ${\cal T}$.
Details on the convergence of our numerical results are provided in Appendix~\ref{App:convergence}.
%
%........................................................................................................................................%
\subsection{Entanglement entropy}
\label{ent:sec}
\subsubsection{Dynamics with unitary and measurement parts}
%

%%%%%%%%%%%%%%%%%%%%%%%%%%%%%%%%%%%%%%%%%%%%%%%%%%%%%%%%%%%%%%%%%%%%%%%%%%%%%%%%%%%%%%%%%%%%%%%%%%%%%%%%%%%%%%%%%%%%%%%%%%%%%%
\begin{figure*}[!t]
  \includegraphics[width=\textwidth]{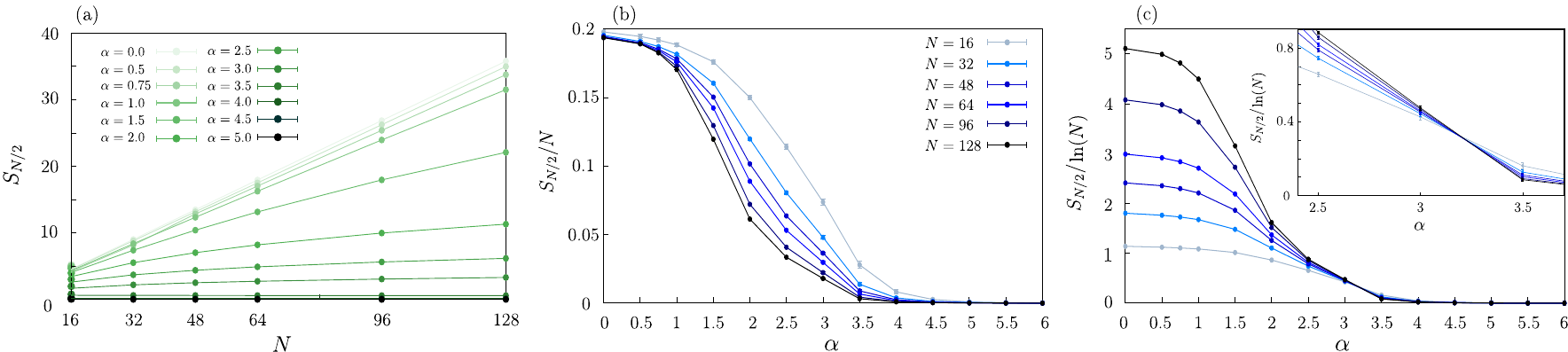}
  \caption{Behavior of the average long-time entanglement entropy $S_{N/2}$ for a system of free fermions,
    governed by the interplay between the Kitaev-Hamiltonian dynamics and the long-range monitoring.
    (a) The entanglement entropy $S_{N/2}$ versus $N$, for different values of $\alpha$
    (increasing $\alpha$ corresponds to a darker color code, as indicated in the legend).
    (b) $S_{N/2}$ divided by $N$ versus $\alpha$, for different sizes
    (increasing $N$ corresponds to darker markers).
    (c) $S_{N/2}$ divided by $\ln N$ versus $\alpha$, for different sizes [same sizes and markers as in panel (b)].
    The inset is a magnification of the same data around $\alpha = 3$.
    We fix $\gamma=0.1$, $J=1$, and $h\!: 100 \to 0.5$.
    Numerical parameters: $\Delta t = 5\cdot 10^{-3}$, $N_{\rm r} \geq 48$,
    errorbars as in Appendix~\ref{App:convergence}.}
  \label{entropie:fig}
\end{figure*}
%%%%%%%%%%%%%%%%%%%%%%%%%%%%%%%%%%%%%%%%%%%%%%%%%%%%%%%%%%%%%%%%%%%%%%%%%%%%%%%%%%%%%%%%%%%%%%%%%%%%%%%%%%%%%%%%%%%%%%%%%%%%%%%

We specifically address the behavior of the average asymptotic entanglement entropy [see Eq.~\eqref{Eq:S_inf}]
as a function of the system size $N$ and of the power-law exponent $\alpha$ for the measurement operator. 
We first consider a free-fermionic system described by the Kitaev Hamiltonian~\eqref{hami:eqn},
and continuously monitored through the long-range operators~\eqref{li:eqn}. 

In Fig.~\ref{entropie:fig}(a) we show $S_{N/2}$ versus the system size $N$,
for different values of $\alpha$ (color gradient).
We notice that, for $\alpha \lesssim 1$, it exhibits a volume-law scaling (i.e., it grows linearly with $N$).
When increasing the power-law exponent, the curves bend to eventually show a flat profile, for very large $\alpha$. 
This behavior suggest the emergence of a volume-law behavior in the long-range regime ($\alpha < 1$) that,
after a crossover for intermediate values of $\alpha$, turns into an area-law behavior
at short-range monitoring ($\alpha \gg 1$).

This can be appreciated more clearly in Fig.~\ref{entropie:fig}(b), where we plot the normalized
asymptotic entanglement $S_{N/2}/N$ versus the power-law exponent $\alpha$. 
As expected, for $\alpha \lesssim 1$ the curves for different values of $N$ collapse to a finite value,
evidencing a linear scaling with $N$. On the other hand, for $\alpha \gtrsim 3.5$,
the curves approach the zero value, thus signaling the onset of a regime where the dependence
of $S_l$ with $N$ is very weak, if not absent (meaning area-law behavior).
In the intermediate regime $1 \lesssim \alpha \lesssim 3.5$, we also observe a less-than-linear dependence
on the system size, which is more difficult to characterize properly.

Further insight on the sublinear region ($\alpha \gtrsim 1$) can be obtained after rescaling the entropy by $\ln(N)$,
as in Fig.~\ref{entropie:fig}(c).
In particular, looking at the inset, the curves for different system sizes exhibit a crossing at $\alpha \sim 3.2$. 
This should correspond to a value marking the transition between a more-than-logarithmic
and a sublogarithmic (most probably area-law) dependence with $N$.
At this point we should note that, since for $\alpha > 2$ the measurement operators have a short-range character,
one cannot rule out the possibility to have a further transition in the intermediate region, from a power-law (sublinear)
to a logarithmic scaling, before ending up into an area-law region at $\alpha \gtrsim 3.2$.
Although hardly visible from our numerical data, the possible occurrence of a logarithmic scaling could be
of the same kind of those emerging in free-fermionic systems in the presence
of local monitoring~\cite{Alberton2021, Buchhold2021, Szynieszewski2022, Poboiko2023}.

Summarizing, we can locate two special points $\alpha^\star_1$ and $\alpha^\star_2$ separating three
regions with qualitatively different behaviors in the entropy scaling with $N$
(increasing $\alpha$, we have volume-law, intermediate subvolume, and area-law scalings of $S_l$ with $N$).
To the best of our numerics, for $\gamma = 0.1$ (corresponding to the data reported in Fig.~\ref{entropie:fig}),
the turning points correspond to $0.5 \lesssim \alpha^\star_1 \lesssim 1$ and $\alpha^\star_2 \sim 3.2$.
While the position of $\alpha^\star_1$ is quite robust when changing the Hamiltonian parameters,
this seems not to be the case for $\alpha^\star_2$. 
In fact, we have performed simulations for other values of $\gamma$ (see, e.g.,
the data for $\gamma=0.5$ in Appendix~\ref{0-5:sec}) and
found that, while the three above regimes (volume-law, intermediate crossover, and area-law) are still present,
the transition point from the intermediate to the area-law behavior %($\alpha^\star_2$)
moves to different values of the power-law exponent (namely, $\alpha^\star_2$ decreases with increasing $\gamma$).
On the opposite hand, we always find $0.5 \lesssim \alpha^\star_1 \lesssim 1$.

%.............................................................................................................................................%
\subsubsection{Measurement-only dynamics}
\begin{figure}
  \includegraphics[width=0.45\textwidth]{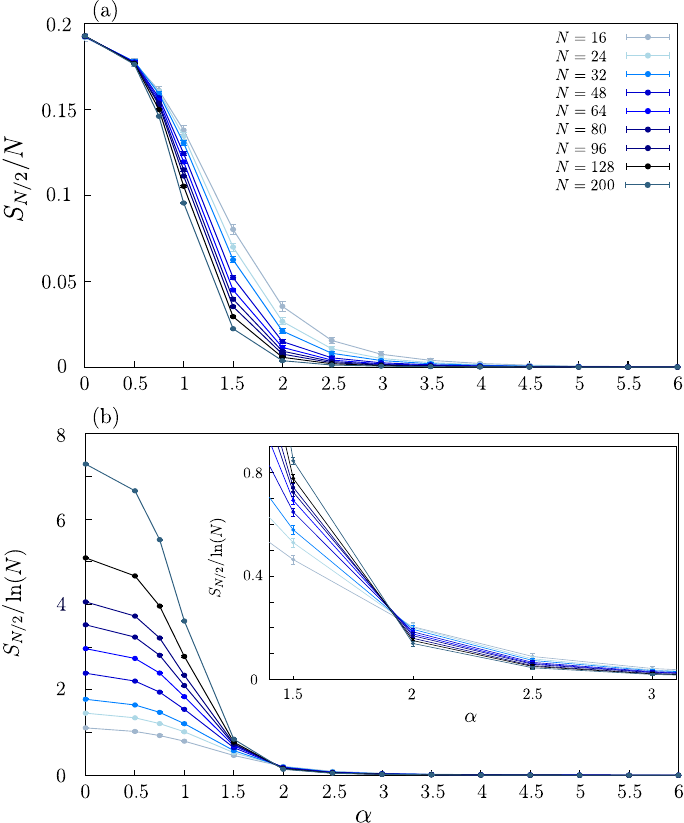}
  \caption{Average long-time entanglement entropy for the case of measuring-only dynamics
    (no Hamiltonian, $J \!=\! h = 0$).
    (a)~$S_{N/2}/N$ versus $\alpha$ for different system sizes.
    (b) $S_{N/2}/\ln N$ versus $\alpha$ for different system sizes.
    The inset is a magnification of the same data around $\alpha = 2$.
    Numerical parameters: $\gamma\Delta t = 5\cdot 10^{-4}$, $N_{\rm r} \geq 48$,
    errorbars as in Appendix~\ref{App:convergence}, same initial state as in Fig.~\ref{entropie:fig}.}
  \label{entropie_nh:fig}
\end{figure}

We now switch to the study of a measurement-only dynamics, i.e., for the case without a Hamiltonian providing
a unitary part in the dynamics ($J = h = 0$).
The plot of $S_{N/2}/N$ versus $\alpha$ is provided in Fig.~\ref{entropie_nh:fig}(a)
[corresponding to Fig.~\ref{entropie:fig}(b) for the case with Hamiltonian], and $S_{N/2}/\ln N$ versus $\alpha$
can be found in Fig.~\ref{entropie_nh:fig}(b) [corresponding to Fig.~\ref{entropie:fig}(c)
  for the case with Hamiltonian].

We notice that the behavior in the small $\alpha$-dynamics is quite stable and, in particular,
it exhibits a volume-law scaling with $L$, for $\alpha \lesssim 1$.
This suggests that the transition point at $0.5 \lesssim \alpha^\star_1 \lesssim 1$ should not depend on the presence
of a Hamiltonian and that it is a property of the measurement operators only. 
There is still an intermediate region featuring a subvolume scaling that vanishes at $\alpha^\star_2 \sim 1.9$,
corresponding to the intersection point of the curves $S_{N/2}/\ln N$ [c.f., the crossing point for the curves in
the inset of Fig.~\ref{entropie_nh:fig}(b)].

%%%%%%%%%%%%%%%%%%%%%%%%%%%%%%%%%%%%%%%%%%%%%%%%%%%%%%%%%%%%%%%%%%%%%%%%%%%%%%%%%%%%%%%%%%%%%%%%%%%%%%%%%%%%%%%%%%%%%%%%%%%%%%%
%%%%%%%%%%%%%%%%%%%%%%%%%%%%%%%%%%%%%%%%%%%%%%%%%%%%%%%%%%%%%%%%%%%%%%%%%%%%%%%%%%%%%%%%%%%%%%%%%%%%%%%%%%%%%%%%%%%%%%%%%%%%%%%

The fact that $\alpha_1^*$ appears to be independent of the system parameters suggests us a comparison with other
long-range systems.
From one side, it is known that long-range Hermitian Hamiltonians in one dimension behave mean-field
for $N\to\infty$ for $\alpha < 1$, short-range for $\alpha>2$, and for $1<\alpha<2$ there is an intermediate
regime where the excited states of the system can break a symmetry, but in a non-mean-field way~\cite{Ruffo,PhysRevLett.120.130601}.
In the case without Hamiltonian, the dynamics is provided by a long-range noisy (pseudo) Hamiltonian in imaginary time
[see Eq.~\eqref{stoca:eqn}], and it is interesting that the transition points of the
dynamics ($0.5 \lesssim \alpha_1^*\lesssim 1$ and $\alpha_2^*\sim 1.9$) approximately coincide with those of the unitary dynamics.
We also recall that, at least in one case~\cite{PhysRevB.106.224308}, $\alpha=1$ is the threshold below which
  a mean-field description is exact for $N\to\infty$ in a Lindblad dynamics in one dimension with long-range Lindbladians.
  
  We conclude the section with a remark on the behavior of $\alpha_2^*$. Considering that the measurement-only case corresponds to the limit of infinite $\gamma$ (more precisely $\gamma\gg h,\,J$), we find that $\alpha_2^*$ decreases with $\gamma$, as we can see in table~\ref{tab:tab}.
\begin{table}
  \begin{tabular}{|c|c|}
    \hline
    $\gamma$&$\alpha_2^*$\\
    \hline
    $0.1$&$\sim 3.5$\\
    $0.5$&$\sim 2.4$\\
    $\infty$ (measurement only) & $\sim 1.9$\\
    \hline
  \end{tabular}
	\caption{Values of $\alpha_2^*$ versus $\gamma$. ($h=0.5$, $J=1$ for the first two rows). More details on the case $\gamma=0.5$ in Appendix~\ref{0-5:sec}.}\label{tab:tab}
\end{table}
%..............................................................................................................................................%
%
\subsection{Expectation probability distribution of the measurement operators}
\label{dist:sec}

%%%%%%%%%%%%%%%%%%%%%%%%%%%%%%%%%%%%%%%%%%%%%%%%%%%%%%%%%%%%%%%%%%%%%%%%%%%%%%%%%%%%%%%%%%%%%%%%%%%%%%%%%%%%%%%%%%%%%%%%%%%%%%%
\begin{figure}
  \includegraphics[width=0.49\textwidth]{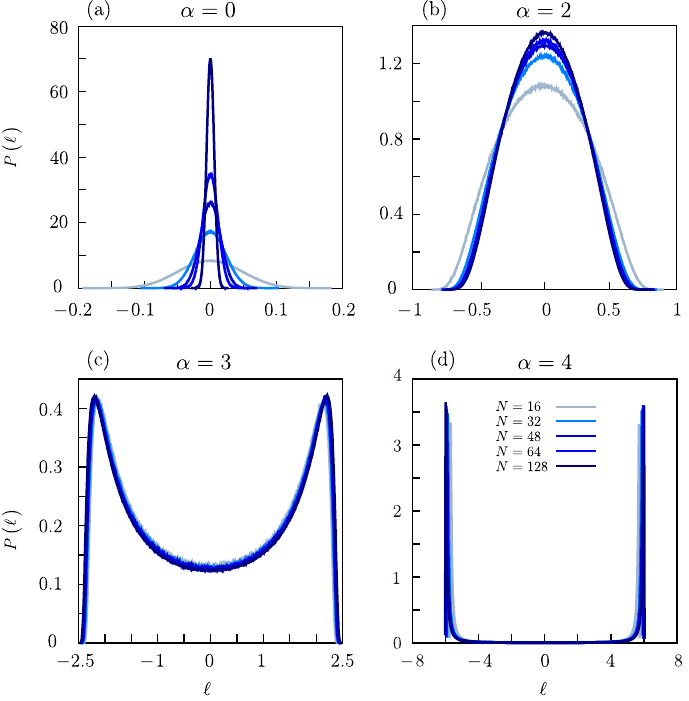}
  \caption{Probability distributions of the expectations $\ell$ of measurement operators,
    over the sites ($j$) and the discretized time ($t$), for various system sizes (see legend),
    The four panels correspond to different values of $\alpha = 0$, $2$, $3$, $4$.
    Here we fix $\gamma=0.1$, $J=1$, and $h: 100 \to 0.5$. Averages over one single quantum trajectory, evolution up to $\mathcal{T}=10^4$, other numerical parameters as in Fig.~\ref{entropie:fig}.}
  \label{Fig:disth}
\end{figure}
%%%%%%%%%%%%%%%%%%%%%%%%%%%%%%%%%%%%%%%%%%%%%%%%%%%%%%%%%%%%%%%%%%%%%%%%%%%%%%%%%%%%%%%%%%%%%%%%%%%%%%%%%%%%%%%%%%%%%%%%%%%%%%%

We now consider the statistics of the expectations of the measurement operators, a quantity that is
experimentally more relevant, being provided by the expectation values of a physically observable operator.
Recent studies have pointed out that, for local measurements, the different properties of this distribution
or related quantities may be connected to the entanglement transitions~\cite{Ladewig2022, Tirrito2022, Turkeshi2023}.

Operatively, we consider a single quantum trajectory, evolve it up to a time ${\cal T}$
  with a given discretization time $\Delta t$, and evaluate all the expectations $\braket{\hat \ell_j}_{t_n}$,
  for the different discrete times $t_n = n \, \Delta t$, ($n = 1, \ldots, {\cal T}/\Delta t$),
  and the different sites $j=1, \ldots, N$. Then we arrange these data into a normalized histogram.
  This is the distribution of the expectations of the measurement operators and we call it $P(\ell)$.

%%%%%%%%%%%%%%%%%%%%%%%%%%%%%%%%%%%%%%%%%%%%%%%%%%%%%%%%%%%%%%%%%%%%%%%%%%%%%%%%%%%%%%%%%%%%%%%%%%%%%%%%%%%%%%%%%%%%%%%%%%%%%%%
\begin{figure}
  \includegraphics[width=0.37\textwidth]{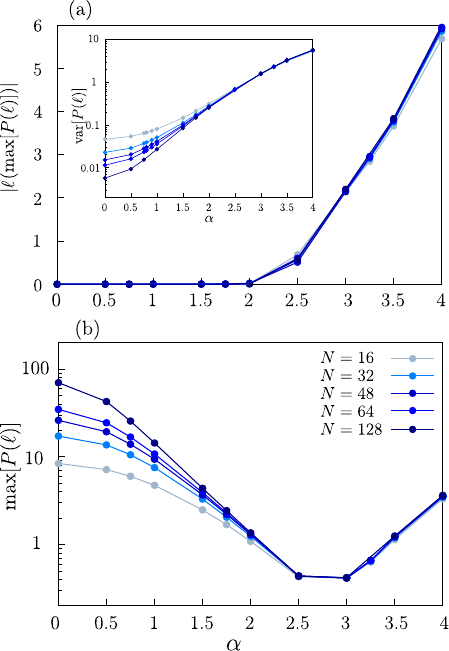}
  \caption{(a) Absolute value of the position of the maximum of the distributions of the $\ell$ versus $\alpha$,
    for different values of $N$. The inset shows the variance of the distribution versus $\alpha$, for different sizes.
    (b) Absolute maximum of the distribution versus $\alpha$ for different sizes. Same parameters as in Fig.~\ref{Fig:disth}.}
  \label{Fig:dist_max}
\end{figure}
%%%%%%%%%%%%%%%%%%%%%%%%%%%%%%%%%%%%%%%%%%%%%%%%%%%%%%%%%%%%%%%%%%%%%%%%%%%%%%%%%%%%%%%%%%%%%%%%%%%%%%%%%%%%%%%%%%%%%%%%%%%%%%%

In Fig.~\ref{Fig:disth} we show, for the dynamics of the monitored Kitaev chain with $\gamma=0.1$,
the histograms of the probability of $\ell$
for $\alpha = 0$ (a), $\alpha = 2$ (b), $\alpha = 3$ (c), and $\alpha = 4$ (d).
The various curves in each panel are for different system sizes (color gradient -- see legend).
The distributions for $\alpha>1$ tend to a limit for increasing system size,
while for $\alpha\leq 1$ there is a rescaling (i.e., in the latter case, the distributions converge
  to a limit, if appropriately rescaled).

It is evident that the shape of such distribution exhibits a crossover from a unimodal to a bimodal character,
depending on the value of $\alpha$. As we have already emphasized, this is reminiscent of bifurcations
in nonlinear classical driven-dissipative dynamical systems~\cite{strogatz:book, cross:book},
where one stationary point splits into two. Here we have also quantum fluctuations and classical noise,
so instead of having stationary points, we have maxima that move to be one (unimodal) to be two (bimodal).

To locate the turning point $\bar{\alpha}$, in the main panel of Fig.~\ref{Fig:dist_max}(a) we plot the absolute value
of the position of the maximum $|\ell(\,\text{max}[P( \ell)]\,)|$ versus $\alpha$. The latter starts deviating from zero
at $\bar{\alpha} \gtrsim 2$, that is far from both the crossover points we identified from the entanglement dynamics
($0.5 \lesssim \alpha^\star_1 \lesssim 1$ and $\alpha^\star_2 \sim 3.2$, for $\gamma=0.1$). 
The inset of Fig.~\ref{Fig:dist_max}(a) shows the variance of the distribution (logarithmic scale on the y-axis). 
For $\alpha \lesssim 1$ the variance is size dependent, meaning that the distribution shrinks when increasing $N$. 
This dependence is still present for $1 < \alpha \lesssim 2$, but it seems to disappear for larger system sizes. 
For $\alpha >2$, no size dependence is observed and, according to the bimodal character of the distribution, the variance becomes sensitively larger. 

Different information can be extracted by looking at the value of the absolute maximum of the distribution,
$\text{max}\big[P(\ell)\big]$, shown in Fig.~\ref{Fig:dist_max}(b) (logarithmic scale on the y-axis). 
The first observation is that, in accordance with the variance behavior, the absolute maximum exhibits a strong size dependence for any $\alpha \lesssim 1$. 
This size dependence is still present at small $N$ for $1 < \alpha \lesssim 2$ to eventually disappear for larger power-law exponents.  
Then we notice that $\text{max}\big[P(\ell)\big]$ shows a non-monotonic behavior in $\alpha$. The absolute minimum
occurs not far from the $\alpha^\star_2$ value at which we observed the transition to the area-law regime of the entanglement entropy.
Since we do not have any theoretical insight, we do not make any direct connection between the two transitions.

We finally comment that a different scenario emerges for the measurement-only dynamics.
In fact, in this case we observe the transition from unimodal to bimodal character at $\bar{\alpha} \sim 1$.
As discussed in Sec.~\ref{ent:sec}, this value corresponds to that of $\alpha^\star_1$,
at which we observe the crossover of the entanglement entropy
from the volume-law to the subvolume-law phase (c.f. Fig.~\ref{entropie_nh:fig}).
This result is consistent with the hypothesis that the interplay with the Hamiltonian can generate an intermediate region
displaying more complex features.
No clear information can be extracted by the analysis of the maxima nor of the moments of the distribution (e.g., the variance).

%.......................................................................................................................................%

\section{Conclusion}\label{conc:sec}
We have studied the dynamics of the entanglement entropy of a fermionic Kitaev chain undergoing a quantum state diffusion evolution,
as a result of a continuous measurement process generated by two-point power-law decaying operators.
This dynamics preserves the Gaussianity of the state, allowing us to simulate systems up to few hundreds of sites.

First, we focused on the asymptotic entanglement entropy, averaged over the different stochastic measurement processes,
both as a function of the system size and of the power-law measurement exponent $\alpha$. 
We found three regimes: For $\alpha < \alpha^\star_1$ (with $0.5 \lesssim \alpha^\star_1 \lesssim 1$),
the entanglement scales linearly with the system size $N$, that is, as a volume-law; on the opposite hand,
for $\alpha > \alpha^\star_2$ (with $\alpha^\star_2$ dependent on the parameters of the system). it exhibits
a sublogarithmic (probably area-law) scaling.

A similar behavior emerges when considering the measurement-only dynamics. 
In this case, the transition from volume-law to the non-trivial phase roughly occurs at the same
value of $0.5 \lesssim \alpha^\star_1 \lesssim 1$ observed for the full Hamiltonian and measurement-induced evolution,
suggesting that this transition is an effect of the measurement process only.
The other transition point at $\alpha^\star_2$, from the subvolume to the area-law phase,
shifts to a smaller power-law exponent. 
These findings suggest a comparison with the case of one-dimensional long-range Hamiltonians,
where also two values of $\alpha$ marking a dynamical transition are present:
The investigation of a possible connection between the unitary case and our non-Hermitian dynamics
may be the focus of future research.

Second, we considered the expectation probability distribution of the measurement operators.
For both the cases of dynamics with and without the Hamiltonian, we have seen that such distribution exhibits
a transition from a unimodal to a bimodal behavior, when increasing $\alpha$ above a given threshold $\bar{\alpha}$. 
However, while for the measurement-only dynamics this transition occurs in correspondence of the $\alpha^\star_1$
at which the entanglement entropy exhibits a transition from volume to subvolume scaling,
in the additional presence of the Kitaev Hamiltonian this correspondence disappears. The absolute maximum of the distribution,
however, behaves non-monotonically in $\alpha$ and exhibits a minimum occurring at a power-law exponent that is
compatible with the transition from the subvolume to the area-law phase.
Nevertheless, this phenomenon is very interesting in itself, being a quantum analog of the bifurcations
occurring in classical driven-dissipative dynamical systems. For that reason we dub it a ``quantum bifurcation''.

In view of the apparently large finite-size effects, to have a confirmation of the stability of the different
system behaviors with $\alpha$, one could look at other quantities as the mutual information or the correlation
functions.
It would be also tempting to investigate the dependence of these results on the specific unraveling.
For example, one can check whether the $\alpha^\star_1$ threshold is robust to the stochastic process
chosen to simulate the Lindblad master equation, i.e., whether it is a property of the operator itself,
as discussed in~\cite{PhysRevB.106.224308}. 
Moreover the effects of long-range measurement operators can be tested in others systems,
as for quantum circuits~\cite{Vasseur2021}.

From an experimental perspective, it is important to investigate the connection between the transition
of the entanglement entropy and the quantum bifurcation of the distribution.

Before concluding, we mention that a remarkably similar phenomenology has been observed in~\cite{Chiriaco2023b}, where a system of monitored two coupled chains of free fermions is considered. 
In this work it is shown that it is possible to induce non-Markovian effects on one of the two chains, referred as the system by performing Markovian measurements on the other one, referred as the bath. 
%This non-Markovianity is reflected in the entanglement dynamics that exhibits three different regimes: an area law scaling, a mixed scaling and a volume law one. 
This non-Markovianity is reflected in the entanglement dynamics that exhibits three different regimes: An area law scaling, a logarithmic scaling and a mixed (logarithmic-volume) scaling. Although it could be interesting to investigate the connection between this non-Markovianity and our non-locality, we leave it to future studies.

\begin{figure}
  \includegraphics[width=0.38\textwidth]{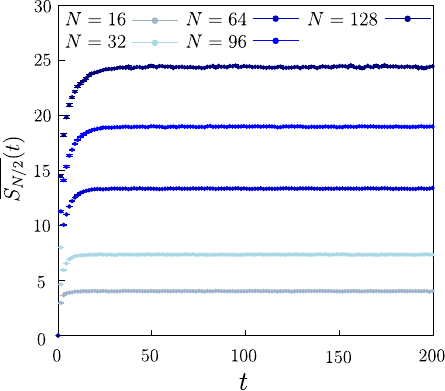}
  \caption{Behavior of the entanglement entropy in time for different system sizes (color scale), $\gamma = 0.1$, $h = 0.5$ and $\alpha = 2$. After a transient, the entanglement entropy saturates to a value that depends on the system size. }
  \label{Fig:ham_vs_t}
\end{figure}
%%%%%%%%%%%%%%%%%%%%%%%%%%%%%%%%%%%%%%%%%%%%%%%%%%%%%%%%%%%%%%%%%%%%%%%%%%%%%%%%%%%%%%%%%%%%%%%%%%%%%%%%%%%%%%%%%%%%%%%%%%%%%%%

\acknowledgments
We thank V. Alba, G. Chiriac\`o, and J. De Nardis for fruitful discussions.
We acknowledge financial support from PNRR MUR project PE0000023-NQSTI. A.~R. acknowledges computational resources from MUR, PON “Ricerca e Innovazione 2014-2020”,
under Grant No. PIR01 00011 - (I.Bi.S.Co.). A.~R. thanks the ICTP for the warm hospitality received (under ERC Project 101053159 -- RAVE) during the preparation of this work. 
We acknowledge support from the Italian MIUR through PRIN Project No. 2017E44HRF.

%.......................................................................................................................................%

\appendix

%.......................................................................................................................................%
\section{Convergence of the numerical results}
\label{App:convergence}

%%%%%%%%%%%%%%%%%%%%%%%%%%%%%%%%%%%%%%%%%%%%%%%%%%%%%%%%%%%%%%%%%%%%%%%%%%%%%%%%%%%%%%%%%%%%%%%%%%%%%%%%%%%%%%%%%%%%%%%%%%%%%%%

All the results have been derived by fixing as integration step $\Delta t = 5 \times 10^{-3}$. This value has been chosen after a convergence check. 

\begin{figure}[!t]
  \includegraphics[width=0.45\textwidth]{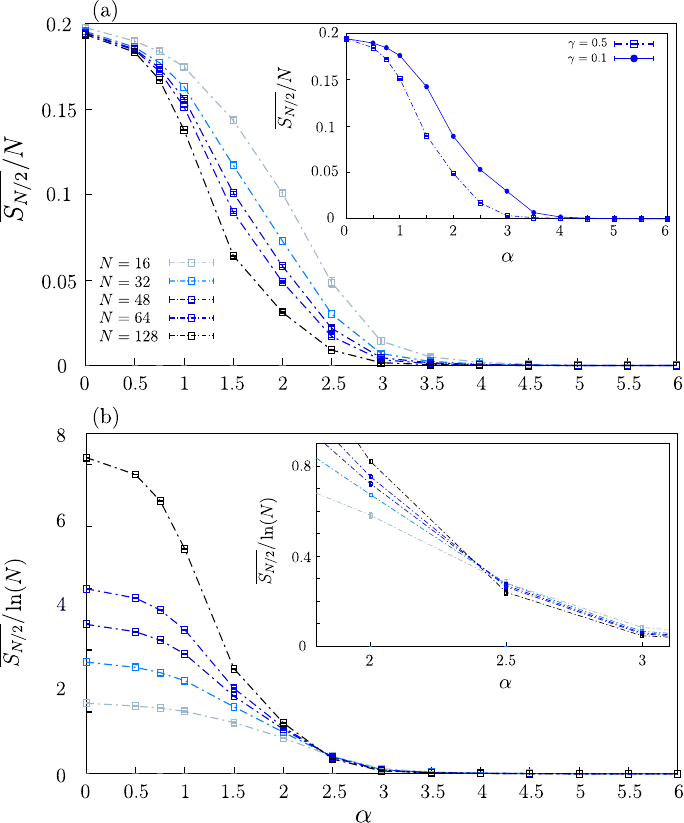}
  \caption{Average long-time entanglement entropy for the case with Hamiltonian ($J = 1$, $h = 0.5$)
    and coupling with the environment $\gamma=0.5$.
    (a) $S_{N/2}/N$ versus $\alpha$ for different system sizes. The inset shows a comparison
    of the curves, for fixed size N=64,
    with $\gamma=0.5$ and $\gamma=0.1$.
    (b) $S_{N/2}/\ln N$ versus $\alpha$ for different system sizes.
    The inset is a magnification of the same data around $\alpha = 2.5$.
    Other numerical parameters and initial state as in Fig.~\ref{entropie:fig}.}
  \label{entropie_0-5:fig}
\end{figure}
{The entanglement entropy is defined in Eq.~\eqref{Eq:S}, we evaluate its average $S_{N/2}$ -- defined in Eq.~\eqref{Eq:S_inf} -- considering a finite averaging time $T$,
which has been chosen so that convergence is attained.}
In Fig.~\ref{Fig:ham_vs_t} we show the characteristic behavior of $_{N/2}$ in time for different system sizes (color scale) and $\alpha = 2$. From this figure it is clear that convergence is reached in reasonable times.

The ensemble average is evaluated over $N_{\rm r} \geq 48$ trajectories. The inequality means that for small system sizes we can easily average over $N_r = O(10^2)$ trajectories, while for larger $N$ the numerical effort required for the simulations does not allow to go beyond $N_r = 48$. 

However, we checked that all the results are consistent inside the error bars $\delta S_{N/2}$, evaluated as 
\begin{equation}
  \delta S_{N/2} = \frac{1}{\sqrt{N_{\rm r}}}\sqrt{\lim_{T\to \infty}\int_{t^*}^T dt' \overline{S_{N/2}^{\,2}(t')}-S_{N/2}^{\,2}}\,.
\end{equation}

{About the numerical implementation, using a FORTRAN code parallelized with Open MPI, for $N=128$, in order to get a time trace of the entanglement entropy on a Intel\textsuperscript{\textregistered} i7 processor with 8 cores of a laptop, for $N_r = 48$s, one needs more or less three days. Because we needed to do these computations for many points in the parameter space, and also for system sizes larger than $N=128$, the use of the cluster mentioned in the acknowledgements was more suitable for us.}

%\dr{[Dire qualcosa anche sulle distribuzioni. Quanta statistica \`e stata presa (fino a che tempi si fa evolvere la
%    singola traiettoria)?]} \ar{\bf [Fino a $\mathcal{T}=10^4$ -- l'ho scritto nella caption delle distribuzioni]}
%
%.........................................................................................................................................%
\section{Case with Hamiltonian and $\gamma = 0.5$}\label{0-5:sec}

%%%%%%%%%%%%%%%%%%%%%%%%%%%%%%%%%%%%%%%%%%%%%%%%%%%%%%%%%%%%%%%%%%%%%%%%%%%%%%%%%%%%%%%%%%%%%%%%%%%%%%%%%%%%%%%%%%%%%%%%%%%%%%%
%%%%%%%%%%%%%%%%%%%%%%%%%%%%%%%%%%%%%%%%%%%%%%%%%%%%%%%%%%%%%%%%%%%%%%%%%%%%%%%%%%%%%%%%%%%%%%%%%%%%%%%%%%%%%%%%%%%%%%%%%%%%%%%

Here we provide results for a case similar to the one considered in Fig.~\ref{entropie:fig},
with the only difference that now $\gamma=0.5$. The corresponding numerical data are shown in Fig.~\ref{entropie_0-5:fig}.
Looking at the plot of $S_{N/2}/N$ versus $\alpha$, we see that the volume-law still persists for small $\alpha$ values,
up to $0.5 \lesssim \alpha_1^*\lesssim 1$ [Fig.~\ref{entropie_0-5:fig}(a)].
From the data in inset at fixed size, notice also that, for $\gamma=0.5$, the entanglement generally drops faster than for $\gamma=0.1$.

On the other hand, the transition from subvolume law to sublogarithm law occurs for a different value of $\alpha_2^*$
($\alpha_2^* \sim 2.4$) as we can see from the crossing of the curves of $S_{N/2}/\ln N$ versus $\alpha$
for different sizes $N$ [Fig.~\ref{entropie_0-5:fig}(inset of panel (b)).]

%\bibliography{biblio.bib}

\end{document}